\documentclass{emulateapj}
\usepackage{epsfig}

\slugcomment{To appear in ApJ Letters}
\shortauthors{Still et al.}
\shorttitle{Quiescent superhumps in V344 Lyr}

\newcommand{\kep}{{\it Kepler}}

\newcommand{\porb}{\ensuremath{P_{\rm orb}}}
\newcommand{\psh}{\ensuremath{P_{\rm sh}}}

\newcommand{\msun}{\ensuremath{M_\sun}}

\begin{document}

\title{Quiescent Superhumps Detected in the Dwarf Nova V344 Lyrae by {\it Kepler}}

\author{
Martin~Still\altaffilmark{1,2},
Steve~B.~Howell\altaffilmark{3},
Matt~A.~Wood\altaffilmark{4},
John~K.~ Cannizzo\altaffilmark{5,6},
Alan~P.~Smale\altaffilmark{6},
} 
\altaffiltext{1}{NASA Ames Research Center, Moffett Field, CA 94035}
\altaffiltext{2}{Bay Area Environmental Research Inst., Inc., 560 Third St. W, Sonoma, CA 95476}
\altaffiltext{3}{National Optical Astronomy Observatory, Tucson, AZ 85719}
\altaffiltext{4}{Department of Physics and Space Sciences, Florida Institute of Technology, 150 W. University Blvd., Melbourne, FL 32901}
\altaffiltext{5}{CRESST/Joint Center for Astrophysics, University of Maryland, Baltimore County, Baltimore, MD 21250}
\altaffiltext{6}{NASA/Goddard Space Flight Center, Greenbelt, MD 20771}

\keywords{stars: novae, cataclysmic variables -- stars: dwarf novae -- stars: individual: V344 Lyr -- stars: white dwarfs}

\begin{abstract}
The timing capabilities and sensitivity of {\it Kepler}, NASA's observatory to find Earth-sized planets within the habitable zone of stars, are well matched to the timescales and amplitudes of accretion disk variability in cataclysmic variables. This instrumental combination provides an unprecedented opportunity to test and refine stellar accretion paradigms with high-precision, uniform data, containing none of the diurnal or season gaps that limit ground-based observations. We present a 3-month, 1 minute cadence {\it Kepler} light curve of V344 Lyr, a faint, little-studied dwarf nova within the {\it Kepler} field. The light curve samples V344 Lyr during five full normal outbursts and one superoutburst. Surprisingly, the superhumps found during superoutburst continue to be detected during the following quiescent state and normal outburst. The fractional excess of superhump period over the presumed orbital period suggests a relatively high binary mass ratio in a system where the radius of the accretion disk must vary by less than 2\% in order to maintain tidal precession throughout the extended episode of superhumping. Disk radius is less restricted if the quiescent signal identified tentatively as the orbital period is a negative superhump, generated by a retrograde-precessing accretion disk, tilted with respect to the binary orbital plane.
\end{abstract}

\section{Introduction}

NASA's Discovery mission {\it Kepler} (Borucki et al. 2010; Haas et al. 2010) was launched on 2009 March 6  upon a focused mission to detect and characterize terrestrial planets within the habitable zone of stars. The {\it Kepler} instrument is a 0.95-m Schmidt camera with a 116 square degree field of view. The single photometric bandpass is broad, 95\% of the response resides between $\lambda\lambda$\,4,230--8,970\AA\ (Koch et al. 2010). Instrument commissioning performance is described by Caldwell et al. (2010). Additional to planet hunting, the photometric accuracy, cadence and uninterrupted integration of the same field over a multi-year mission provides unique opportunities to monitor sources of astrophysical interest.

{\it Kepler}Õs field of view contains ten known cataclysmic variables (CVs), listed in Table~\ref{tab1} with celestial coordinates, magnitude range, orbital period and sub-class, where known. These sources represent a specific and dynamic stage in binary star evolution. The primary (more massive) stellar component is a white dwarf. In the past its progenitor star evolved off the main sequence and its expanding outer envelope engulfed a companion star. In a short-lived common-envelope episode, angular momentum was removed efficiently from the binary, leaving a close pair with separation a few stellar radii and orbital period a few hours (see Warner 1995 for a review). Stable angular momentum loss continues slowly after the common envelope phase through magnetic braking and gravitational radiation (Paczy\'{n}ski 1967; Rappaport, Verbunt \& Joss 1983). Through these angular momentum transfer mechanisms, the companion star's Roche lobe descends upon and makes contact with the companion. Contact initiates ballistic mass transfer from the companion star to the white dwarf through the inner Lagrangian point, at which time the object is classed a CV. Mass transfer is stable and continues generally until the companion becomes degenerate. Thus, CVs have broad importance for constraining models of stellar evolution, binary evolution and the chemical enrichment of the Galaxy.

\begin{figure*}
\epsfig{file=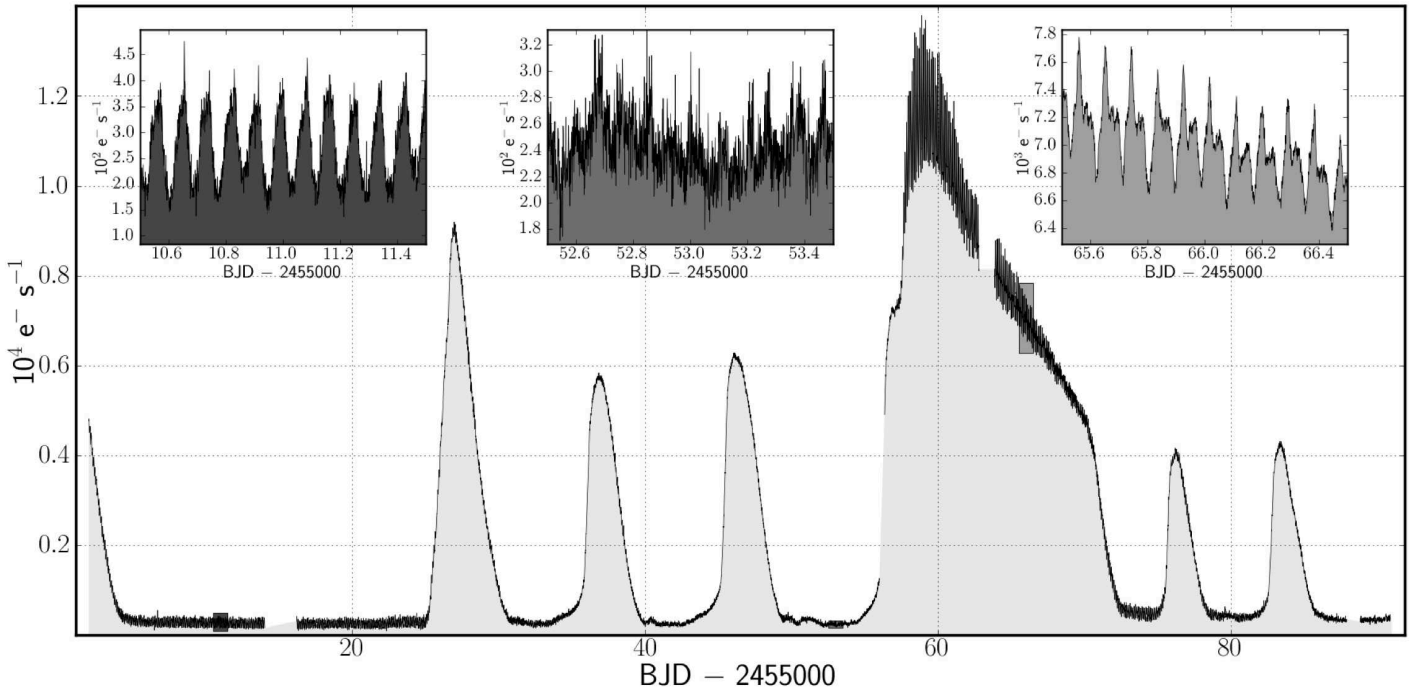, angle=270,width=7.3in,bbllx=100,bblly=40,bburx=490,bbury=790}
\caption{The quarter 2 {\it Kepler} light curve of V344 Lyr. The insets are magnifications of the three boxes superimposed over the light curve at discrete epochs. From left to right, insets contain a quiescent light curve dominated by photometric variability on, or close to, the putative orbital period, a quiescent light curve with no coherent orbital signal, and a superoutburst light curve dominated by double-peaked superhumps.}
\label{fig1}
\end{figure*}

 The CV phase has another important feature. Ballistic material leaving the companion star cannot fall onto the white dwarf without losing angular momentum. In the absence of a powerful magnetic field from the white dwarf, material does this through the shearing layers of a circumstellar accretion disk (Shakura \& Sunyaev 1973). Such disks occur in nature over a range of scales, from planet-building disks around protostars to those that feed the distant quasars. However, CV disks are unique in that their optical light dominates over both stellar companions, and systems are typically nearby, within a kpc. The mass transfer rate through the disk is often modulated by a viscously driven limit cycle (Smak 1984), resulting in intense, short-lived optical outbursts. Because of the observational detail available, CVs are primary targets for monitoring accretion disk behavior and testing accretion disk theory.

\begin{table*}
\label{tab1}
\caption{Identified cataclysmic variables in the Kepler field. All data are extracted from the online catalog of Downes et al. (2001), except where noted.}
\begin{center}
\begin{tabular}{rccccrl}\hline\hline
Name & KepID\footnote{Target identification number from the Kepler Input Catalog.} & $\alpha$ (J2000) & $\delta$ (J2000) & $m_{\mbox{\footnotesize v}}$ & $P_{\mbox{\footnotesize orb}}$ (h)\footnote{Orbital period.} & Sub-class\footnote{See Warner (1995) for a definition of CV sub-classes.}\\
\hline
V344 Lyr & 7659570 & 18 44 39.17	& +43 22 28.2 & 13.8--20.0 & & SU UMa\\
V358 Lyr & & 18 59 32.95 & +42 24 12.2 & 16.0--20.0 & & WZ Sge\\
V447 Lyr	& 8415928 & 19 00 19.92 & +44 27 44.9 & 17.2--18.5 & & dwarf nova\\
MV Lyr & 8153411 & 19 07 16.29 & +44 01 07.7 & 12.1--17.7 & 3.18 & nova-like\\
V452 Lyr & 7742289 & 19 10 26.32 & +43:28:55.2 & 17.6--18.5 & & dwarf nova\\
V585 Lyr & & 19 13 58.40 & +40 44 09.0 & 14.9--21.1 & & SU UMa\\
V516 Lyr & 2436450 & 19 20 35.73 & +37 44 52.3 & 18.9--22.2 & & dwarf nova\\
V523 Lyr & & 19 21 07.40 & +37:47:56.5 & 17.7--20.2 & & VY Scl\\
& 8751494 & 19 24 10.82 & +44 59 34.9 & 15.8 & 2.94 & nova-like\footnote{Williams et al. (2010)}\\
V1504 Cyg & 7446357 & 19 28 56.47 & +43 05 37.1 & 13.5--17.4 & 1.67 & SU UMa\\
\hline
\end{tabular}
\end{center}
\end{table*}

\begin{figure*}
\epsfig{file=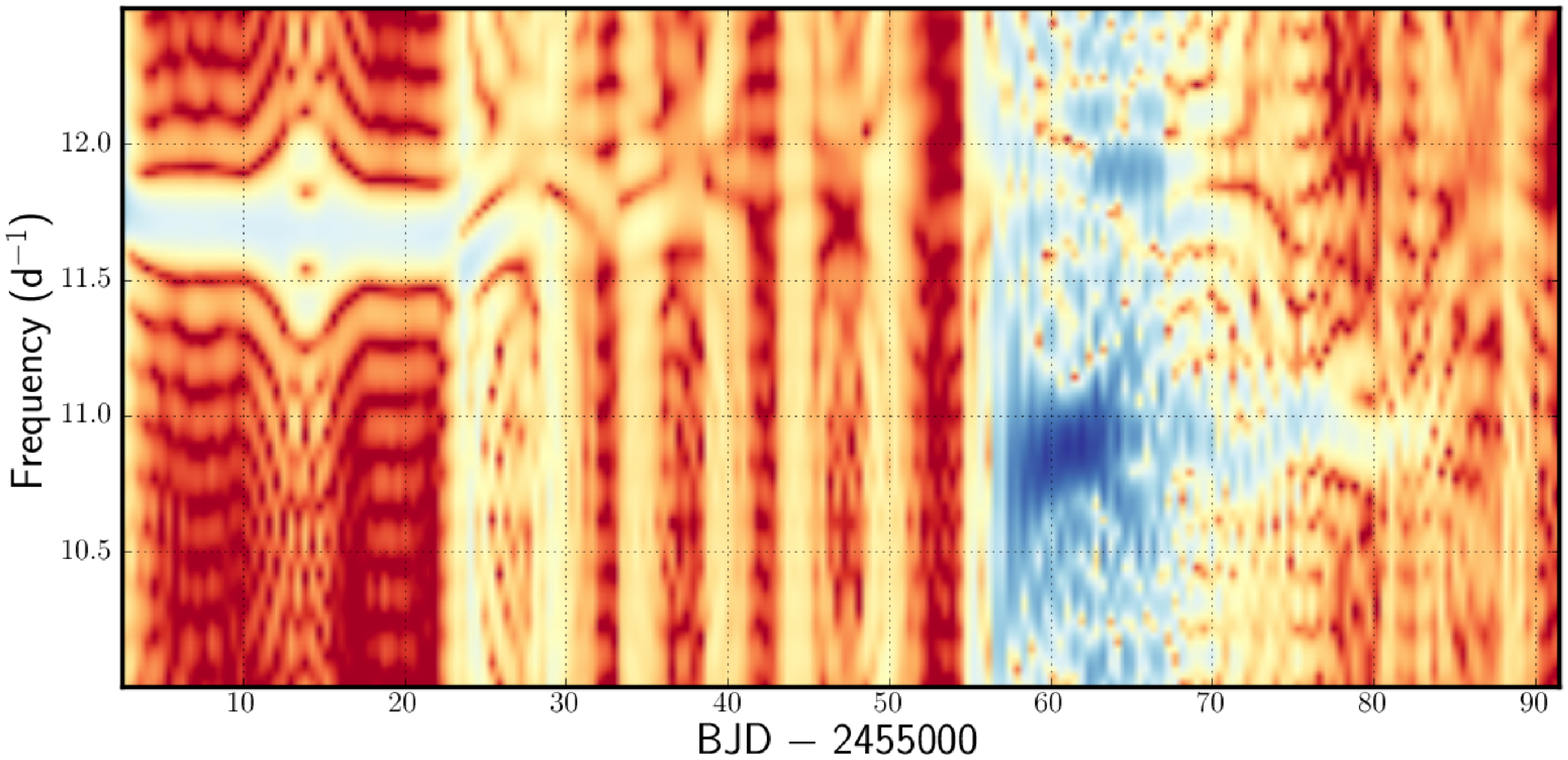,angle=270,width=7.3in,bbllx=100,bblly=40,bburx=490,bbury=790}
\caption{Discrete Fourier Transform power spectra calculated over the duration of Q2 in 5 day slices, stepped by 0.5 days. Red indicates minimum power, blue indicates maximum power. The putative orbital period, \porb, is detected at 11.7 cycles\,d$^{-1}$ but is not coherent, disappearing from the light curve before BJD 2,455,040. The putative superhump period, \psh, is detected at 10.8 cycles\,d$^{-1}$, but is not confined to the superoutburst, and is detected beyond BJD 2,4Ä55,080.}
\label{fig2}
\end{figure*}

Using the unprecedented detail of {\it Kepler} quiescent-to-outburst light curves we can test viscous accretion disk limit cycle models, quantify mass accretion rates, disk viscosity and trace the evolution of cooling and heating waves across the disk. {\it Kepler} will monitor the evolution of superhump periods throughout superoutbursts (Osaki 1989), testing the viscous disk (Shakura \& Sunyaev 1973), mass transfer instability (Osaki 1970), Enhanced Mass Transfer (EMT; Osaki 1985), Thermal-Tidal Instability (TTI; Osaki 1989) and resonant disk (Whitehurst 1988) models directly against unique photometric observations. We can also characterize disk flickering, quantifying its properties and the inferred mass transfer viscosity evolution as a function of luminosity and accretion rate.  Sensitive searches for the spin periods of the white dwarfs may reveal the magnetic nature and strength of the accretor and the truncation of the inner accretion disk due to magnetic pressure. We can also conduct systematic searches for orbital eclipses, yielding the inclination and mass ratio of the binary, greatly increasing the accuracy of accretion models.

V344 Lyr is a dwarf nova which resides in the {\it Kepler} field. It is a faint source, with a quiescent magnitude of V $\simeq$ 19, and therefore has been observed sparsely until now. Optical photometry by Kato (1993) coincided with a bright epoch of V $\simeq$ 14 and revealed the presence of putative 2.1948 $\pm 0.0005$ hour superhumps, indicating that this source is a member of the SU UMa class of dwarf novae. The normal outbursts occurred over a 16 $\pm$ 3 day cycle and the superoutburst recurred on a mean cycle of 110 days (Kato, Poyner \& Kinnunen 2002). Ak et al. (2008) used an orbital period-luminosity relationship to estimate a 619 pc distance to V344 Lyr, although the orbital period has, to date, not been confirmed directly.

Relative to normal outbursts, superoutbursts are a magnitude or more brighter, considerably longer, have a shallower decay profile and reveal superhumps -- photometric oscillations on a period a few percent longer than the binary orbit (see Kato et al. 2009). The main competing models of superoutbursts, TTI and EMT, require the accretion disk to expand in size up to or beyond the radius where the rotation rate of the disk is in a 3:1 resonant state with the orbital period (Whitehurst 1988). Additional flux detected during superoutburst is the result of increased mass transfer through the accretion disk externally driven by the companion star through tidal dissipation. The superhumps are a combination of tidally-driven precession of spiral arms in the outer disk and the reaction to precession of the classical hot spot located where the ballistic accretion stream impacts the disk.

During the first 90 days of observation, {\it Kepler} makes the unusual detection of superhumps within superoutburst, normal outburst and quiescent source states. This behavior is not predicted by the conceptual versions of superoutburst models.

\section{Data Analysis}

{\it Kepler} data are provided by the Science Operations Center to analysts in reduced and calibrated form after being propagated through the standard data reduction pipeline (Jenkins et al. 2010). Only the Simple Aperture Photometry (SAP) light curve provided for V344 Lyr is considered in this paper. The Pre-search Data Conditioning (PDC) light curve, designed to optimize the search for 6 hour duration planet transits, is less well-suited to the accurate representation of high-amplitude photometric variability on timescales $> 10$ days. Consequently data artifacts that are removed in PDC data remain in our {\it Kepler} SAP. 

Artifacts include those caused by rapid attitude tweaks to correct pointing drift and those caused by the loss of fine-pointing. Fine-point loss is most often caused by thruster firings, performed on regular 3 day intervals to desaturate the spacecraft reaction wheels. Each event can affect {\it Kepler} data adversely for several minutes. Further gaps within the time series occur due to the spacecraft entering occasional anomalous safe modes. Data are not taken during such events and recorded data following a safe mode are often correlated with electronic warming over the first few days of operation. Further details of all artifacts and a list of their occurrence times are provided in the {\it Kepler} Data Release Notes \#3\footnote{http://archive.stsci.edu/kepler/documents.html}. In this paper we make no attempt to correct data artifacts and simply remove them from the time series.

The large panel in Figure~\ref{fig1} plots the short cadence (1 min; Gilliland et al. 2010), barycentric-corrected light curve of V344 Lyr\footnote{A Guest Observer Office target with KeplerID: 7659570} from the second quarter (Q2) of {\kep} operation which spans the interval 2009 June 20 to September 16. Data gaps correspond to removed artifacts and monthly data downloads. The quiescent V band magnitude is reported contemporaneously by the American Association of Variable Star Observers to be V $\simeq$ 19. While we make no attempt at a color correction, these data indicate that {\it Kepler} short cadence data achieves 4\% relative photometric accuracy upon V = 19 sources, in agreement with Gilliland et al. (2010). The data reveal seven consecutive dwarf nova outbursts over the quarter of varying magnitude, duration and interval, with a comparatively short duty cycle on the order of 10 days. One outburst is a superoutburst, confirming V334 Lyr as an SU UMa class object, first reported by Kato (1993). 

Discrete Fourier Transforms over two short data samples suggest we are detecting superhumps during superoutburst.  Data extracted from BJD\,2,455,006--2,455,025 reveal a peak at what we will initially refer to as the orbital period, \porb\ = 2.05737 $\pm$ 0.00001 hours, although we argue for an alternative potential origin of this signal in Sec.~\ref{discussion}. The error is a 1$\sigma$ estimate based upon 1,000 Monte Carlo trials where the transform has been recalculated after recasting each point in the photometric curve according to the inverse normal cumulative function and a random number generator. After 1,000 trials all period measurements occurred within the range  2.05733--2.05739 hours. A second epoch was extracted during the superoutburst over range BJD\,2,455,058--2,455,080. While the power spectrum reveals a stronger but less coherent signal compared to the earlier epoch, the best period was determined by identical method to be \psh\ = 2.2023 $\pm$ 0.0001 hours and identified as a superhump. The fractional excess of the superhump period over the orbital period is often characterized by $\epsilon^+ =$ (\psh\ $-$ \porb)/\porb. If we have identified the two signals accurately, The V344 Lyr $\epsilon^+$ = 7.044 $\pm$ 0.005\%. Consequently this binary would occur at the extreme end of the $\epsilon^+$ distribution of the SU UMa class of stars tabulated by Patterson (2005).

Neither the putative orbital nor superhump periods are permanent features of the Q2 light curve. We performed a dynamic Fourier power analysis of V344 Lyr by taking 5 day segments stepped by 0.5 days and calculating Discrete Fourier transforms on each time slice. Power spectra are provided in Figure~\ref{fig2}. The somewhat surprising finding is that the quiescent photometric variability found in the post-superoutburst data at BJD\,2,455,075 does not have the same nature as the quiescent signal measured at BJD\,2,455,010. Post-superoutburst, and throughout the following normal outburst, V344 Lyr continues to be dominated by superhumps. The TTI model struggles generally to predict superhumps within quiescence or normal outbursts and so the {\it Kepler} data provide us with an immediate challenge.

\section{Discussion}
\label{discussion}

\begin{figure}
\epsfig{file=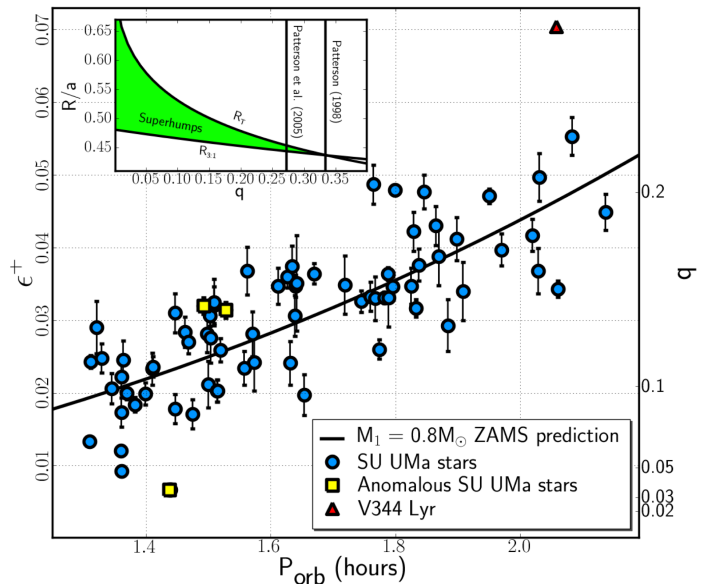, angle=-90, width=3.8in,bbllx=10,bblly=70,bburx=600,bbury=790}
\caption{The measured and $M_1$ = 0.8\,$M_\odot$ zero-age main sequence prediction of the orbital period--superhump period excess relation. V344 Lyr is represented by the red triangle. The inset plots the 3:1 resonance radius, $R_{\rm 3:1}$, and the tidal truncation radius of the disk, $R_{\rm T}$. Disks precessing to the 3:1 resonance must reside between these two limits. Two vertical lines illustrate the mass ratio predicted by the model-dependent $\epsilon^+$--$q$ relation of Patterson (1998) and the empirical relation of Patterson et al. (2005).}
\label{fig3}
\end{figure}

The detection of post-superoutburst superhumps in a CV is rare, but not unique. Two  members of the SU UMa class, V1159 Ori and ER UMa have also revealed superhumps within quiescence or normal outburst (Patterson et al. 1995; Gao et al. 1999). The WZ Sge star EG Cnc has also revealed superhumps up to 90 days after the decay from superoutburst (Patterson et al. 1998). 

Hellier (2001) has argued that quiescent and normal outburst superhumps are possible in extreme mass ratio objects where $q \lesssim 0.07$ ($q = M_2/M_1$; $M_2$ is the companion star mass, $M_1$ is the white dwarf mass). Such a scenario is possible because angular momentum transfer by tidal dissipation is comparatively efficient in this regime; both cooling and heating fronts can propagate even when the outer edge of the accretion disk exceeds the 3:1 resonance radius. Although no confirmed orbital period for this object exists currently, we ask in this section whether V344 Lyr fits this same low-mass ratio profile. 

In order to proceed we must employ a small series of empirical relations. From Patterson (2005) we extract the orbital and superhump periods measured from a sample of SU UMa stars. The distribution of $\epsilon^+$ within this sample is plotted over \porb\ in Figure~\ref{fig3}. Standard SU UMa objects are represented by circle symbols, while the three previous quiescent superhump examples are represented by square symbols. Based upon the empirical fitting of superhump sources with relatively secure measured mass ratios, Patterson (2005) provides the relation $\epsilon^+ = q\left(0.18 + 0.29q\right)$; we provide the predicted $q$ on the right-hand vertical axis of Figure~\ref{fig3}. The curve overlayed upon the distribution is a theoretical $\epsilon^+$--\porb\ prediction using the $\epsilon^+$--$q$ relation and the assumptions that i) $M_1$ = 0.8\,\msun, and ii) the orbital period-companion star density relation and zero-age main sequence (ZAMS) companion star mass-radius relation parameterized from the data of Clemens et al. (1998) are accurate; see Patterson (1998) for a more detailed description. The choice of $M_1$ = 0.8\,\msun\ is precipitated by a good, model-dependent match to the data in Figure~\ref{fig3} and the photometric timing measurements of eclipsing dwarf novae which indicate a more massive white dwarf population in CVs relative to single white dwarfs (e.g. Littlefair et al. 2008). Figure~\ref{fig3} reveals relatively low mass ratios for V1159 Ori, ER UMa and EG Cnc. 

The first of two scenarios we consider is that the high-frequency signal detected in Sec.~2 is the orbital period. In this case, the $\epsilon^+$-predicted mass ratio of V344 Lyr, $q = 0.27$, is much larger than the other quiescent superhumpers. V344 Lyr is also not a good match to the $M_1$ = 0.8\,\msun\ relation; the ZAMS $\epsilon^+$--$q$--\porb\ relation and it's dependent assumptions only remains good if $M_1$ = 0.58\,\msun. This is close to the mean mass of single DA white dwarfs (Bragaglia, Renzini \& Bergeron 1995), but would still clearly be a low-mass outlier within the mass distribution of CVs presented here.

Within the upper-left inset of Figure~\ref{fig3} we plot the 3:1 resonance radius, $R_{\rm 3:1}$, as a function of mass ratio and the tidal truncation radius, $R_{\rm T}$, usually assumed to be 90\% of the mean Roche lobe radius. Disks with outer radii residing between these limits will precess with the 3:1 resonance mechanism. A disk in either the superoutburst, normal outburst and quiescent states will potentially reveal superhumps if it resides within this zone. Based upon the Patterson (2005) relation, this target should undergo precession but only if the outer radius of the disk remains within the range $0.444 < R/a < 0.454$ during all three states, where $a$ is the binary separation. This a relatively contrived parameter space and will provide an interesting challenge for the TTI and EMT models to reproduce the observed outburst phenomenology under such physical restrictions. Patterson (1998) provides an alternative $\epsilon^+$--$q$ relation, $\epsilon^+ = 0.23q / (1 + 0.27q)$, based upon the argument that disk precession rate should be related to secondary star mass and that the most likely sites for tidal excitation are at disk radius $R \approx 0.46a$. This relation is of different form to that obtained empirically and provides a different mass ratio for V344 Lyr of $q = 0.33$.  Under these circumstances, V344 Lyr falls even closer to the upper-$q$ limit where superhumps are possible, $q = 0.34$. This target could also undergo quiescent disk precession but only within a more highly contrived parameter space where the disk radius is very close to the tidal truncation radius throughout the superhump episode, $0.4368 < R/a < 0.4372$.

The second viable scenario we consider is that the shorter of the two signals detected is a negative superhump (e.g. Wood, Thomas \& Simpson 2009). This will occur if the accretion disk becomes tilted relative to the orbital plane, forcing retrograde tidally-driven precession. Negative superhumps are comparatively rare. We note that of the 8 dwarf novae with putative negative superhumps compiled by Wood et al. (2009), the quiescent superhumpers V1159 Ori and ER UMa are contained in that small sample. In such an instance, the binary orbit lies somewhere between our two measured periods. Making the assumption that $M_1 = 0.8$\msun, then $\epsilon^+$ lies in the range 4.6--5.3\%. Consequently we can predict that the negative superhump fractional deficit lies within the range $\epsilon^-$ = 1.7--2.4\% and that $2.086 < \porb < 2.101$ hours. The Patterson (2005) relation infers a mass ratio in this scenario of $0.20 < q < 0.22$ and $0.453 < R/a < 0.478$ during superhump episodes. 

Both scenarios discussed can be tested with time-resolved optical spectroscopy that provides an independent measure of the orbital period. However both scenarios infer that quiescent superhumps are not limited to extreme-low mass ratio CVs.

\section{Conclusion}

With its combination of photometric accuracy, time sampling and fixed field of view, {\it Kepler} opens a previously unexplored window for the detailed monitoring of cataclysmic variables. {\it Kepler} reveals superhumps from V334 Lyr during quiescence, normal outburst and superoutburst. If the two detected signals within the light curve are the binary orbit and disk precession then the inferred mass ratio indicates that the outer disk radius varies by $\lesssim 2\%$ throughout multiple disk state changes, providing very restrictive physical limits for models of superoutburst to replicate. These restrictions become somewhat relaxed if we assume that the detected low-frequency signal is a negative superhump, in which case V344 Lyr is a target of high intrinsic value for accretion disk modeling. A radial velocity-derived orbital period will distinguish between these two scenarios. Ongoing monitoring with {\it Kepler} will also provide further indications of the true nature and coherency of the observed signals. The continued, uninterrupted monitoring by {\it Kepler} of this target, other known CVs, and newly identified objects in the field over the next 3 years or longer will provide a legacy for both the {\it Kepler} mission and CV community.

\acknowledgements 
Funding for this 10th Discovery mission is provided by NASA's Science Mission Directorate. We acknowledge the contributions of the entire {\it Kepler} Team.

\noindent {\it Facilities:} The {\it Kepler} Mission.



\end{document}